# Metamorphosis of Fuzzy Regular Expressions to Fuzzy Automata using the Follow Automata

## Abstract

To deal with system uncertainty, finite automata have been generalized into fuzzy automata. Stamenkovic and Ciric proposed an approach using the position automata for the construction of fuzzy automata from fuzzy regular expressions. There exist multifarious methodologies for the construction of finite automata from regular expressions known as Thompson construction, Antimirov partial derivatives, Glushkov automata and follow automata etc. In this paper, we propose an approach for the conversion of fuzzy regular expressions into fuzzy automata using the concept of follow automata. The number of states of the obtained Fuzzy automata using the proposed approach is lesser than the extant approaches in the literature.

**Keywords:** Fuzzy set, Fuzzy Regular expressions, Fuzzy Automata, Lattice-ordered monoid.

## 1. Introduction

A non-deterministic finite automaton has been generalized into fuzzy automaton. Fuzzy automata has been widely used in diverse applications, such as lexical analysis, learning systems, for describing natural languages, programming language, clinical monitoring, control system, database, learning system, discrete event system, neural network etc. [3, 9]. The study of fuzzy language and automata was introduced for dealing with uncertainty in a system in 1960 by Santos [6]. Various researchers have carried out the preliminary work on fuzzy automata [7, 16-17]. Fuzzy automata can be classified into nondeterministic and deterministic fuzzy finite automata. Fuzzy automata depend on the membership value, which lies between 0 and 1. The membership values in fuzzy automata have been assigned using different types of lattices [5, 11, 19], such as complete residuated lattice, lattice-ordered monoid etc.

Li and Pedrych [19] proposed an approach for the metamorphosis of fuzzy regular expressions to fuzzy automata. An innovative and improved approach was proposed by Stamenkovic and Ciric [3] for the metamorphosis of fuzzy regular expressions into fuzzy automata using the concept of position automata.

Regular expression plays a vital role for string matching, pattern matching, database and searching [8, 13]. Regular expressions can be converted into nondeterministic automata using different techniques, such as Thompson automata [10], position automata [15], follow automata [12], and partial derivative automata [14] etc. Given a regular expression $r$ with length $m$ and $n$ is the number of symbol in $r$. Using Thompson's construction a regular expression can be converted into $\varepsilon$-NFA with 2m states in the worst case. Position automata generate a NFA having n + 1 states from a regular expression. Follow automata [10] is a quotient of position automata, and it is having the number of states less than or equal to $n+1$.

Table 1: Comparison between Position automata and follow automata

| Attribute | Position automata [15] | Follow automata [12] |
|---|---|---|
| Number of states | $n+1$ | $\leq n+1$ |
| Complexity [12] | $O(mn)$ | $O(mn)$ |

These methodologies can be extended for the conversion of fuzzy regular expressions into fuzzy automata. Stamenkovic and Ciric [3] converted the fuzzy regular expressions into fuzzy automata using the concept of position automata. Kumar and Verma [1-2] used the concept of follow position for the conversion of Parallel regular expressions to non-deterministic finite automata. There is scope for producing svelte fuzzy automata using the follow automata for the metamorphosis of fuzzy regular expressions to fuzzy automata.

## 2. Preliminaries

In this section, we describe some basic definition such as lattice-ordered monoid, fuzzy automata and fuzzy regular expressions.

**Definition 2.1:** Let L be a nonempty set, and $\otimes$ is a binary operation $\otimes$, then $(L, \otimes, e)$ is a monoid if it satisfies the following axioms:
   i) Closure Law: $\forall a, b \in L, a \otimes b \in L$
   ii) Associative Law: $\forall a, b, c \in L, (a \otimes b) \otimes c = a \otimes (b \otimes c)$ holds.
   iii) Identity Law: $e \in L, \forall a \in L, a \otimes e = e \otimes a = a$ holds.

**Definition 2.2:** A lattice-ordered monoid [18-20] is an algebraic structure $(L, \wedge, \vee, 0, 1, \otimes, e)$ such that:
   i) $L$ is a lattice with least element 0 and the greatest element 1.
   ii) $(L, \otimes, e)$ is a monoid with identity e.
   iii) $\forall u \in L, u \otimes 0 = 0 \otimes u = 0$
   iv) $\forall u, v, w \in L, u \otimes (v \vee w) = (u \otimes v) \vee (u \otimes w)$, and
   $(u \vee v) \otimes w = u \otimes w \vee v \otimes w$

**Definition 2.3:** $L$ is called quantale [18] if $(L, \wedge, \vee, 0, 1)$ is a complete lattice and satisfy the following condition of infinite distributive laws.

Infinite distributive laws:

$$u \otimes \bigvee_{i \in I} u_i = \bigvee_{i \in I} (u \otimes u_i), \quad \bigvee_{i \in I} u_i \otimes u = \bigvee_{i \in I} (u_i \otimes u),$$

**Definition 2.4:** In lattice-ordered monoid [18-20] $(L, \wedge, \vee, 0, 1, \otimes, e)$, if the identity element $e$ of the monoid $(L, \otimes, e)$ and the greatest element 1 of lattice $(L, \wedge, \vee, 0, 1)$ are coinciding, then $L$ is called an integral lattice-ordered monoid.

**Definition 2.5:** A *fuzzy automaton* FA (Q, Σ, $\delta^Q, \sigma^Q, \tau^Q$) [3-4, 21] over lattice-ordered monoid consists of five tuples, where
1. Q is a finite non-empty set of states.
2. Σ is a finite non-empty set of input alphabets.
3. $\delta^Q : Q \times \Sigma \times Q \to L$ is a fuzzy subset of $Q \times \Sigma \times Q$, called fuzzy transition function or fuzzy transition relation.
4. $\sigma^Q \in L^Q$ is the fuzzy set of initial states.
5. $\tau^Q \in L^Q$ is the fuzzy set of final states.

The fuzzy transition relation $\delta^Q$ takes input as a state from Q, and on reading a symbol from the alphabet, reached any states in Q with membership value. The fuzzy transition function $\delta^Q$ converted into the extended fuzzy transition function $\delta_*^Q$ and mapping $\delta_*^Q : Q \times \Sigma \times Q \to L$ in the following way:

If $q_0, q_1 \in Q$ and $\varepsilon \in \Sigma^*$ is the null word, then,

$$\delta_*^Q(q_0, \varepsilon, q_1) = \begin{cases} 1 & if\ q_0 = q_1, \\ 0 & otherwise, \end{cases}$$

If $q_0, q_1 \in Q$, $u \in \Sigma^*$ and $a \in \Sigma$, then

$$\delta_*^Q(q0, ua, q1) = \bigvee_{a \in \Sigma} \delta_*^Q(q_0, u, q_2) \otimes \delta^Q(q_2, a, q_1).$$

The fuzzy language L(FA) represented by a fuzzy automaton FA = (Q, Σ, $\delta^Q, \sigma^Q, \tau^Q$) can be described using:

$$L(FA)(u) = \bigvee_{q0, q1 \in Q} \sigma^Q(q_0) \otimes \delta^Q(q_0, u, q_1) \otimes \tau^Q(b)$$

**Example 2.2.1:** Consider the fuzzy automata FA = (Q, Σ, $\delta^Q, \sigma^Q, \tau^Q$), where Q = {$q_0, q_1$}, Σ = {0, 1}, fuzzy transition function $\delta^Q$ is shown in fig. 1.

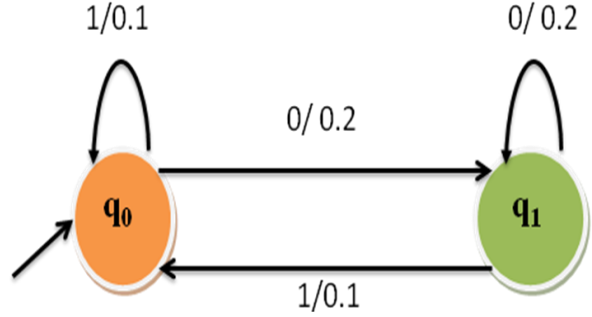

**Fig. 1**: Example of a Fuzzy Automaton.

$\delta^Q(q_0, \varepsilon) = \{\frac{1}{q0}\},$

$\delta^Q(q_0, 10) = \delta^Q_*(q_0, u0, q_1) = V_{0 \in \Sigma}\, \delta^Q_*(q_0, u, q_0) \otimes \delta^Q(q_0, 0, q_1) = \{\frac{0.1}{q_0}\} \cdot \{\frac{0.2}{q1}\},$

$= \{\frac{0.1}{q_1}\}$

$\delta^Q(q_0, 100) = \delta^Q_*(q_0, u0, q_1) = \{\frac{0.1}{q_0}\} \cdot \{\frac{0.2}{q1}\} \cdot \{\frac{0.2}{q1}\},$

$= \{\frac{0.1}{q_1}\}$

We can make use of max–min automata, min–max automata, and max-product automata for determining the language represented by fuzzy automata.

**Definition 2.6:** Let $\Sigma$ be a finite non–empty set of input alphabet. The family *FRE* of fuzzy regular expression [3, 8, 18] can be inductively defined over an alphabet $\Sigma$ using the following rules:

i) $\emptyset \in FRE$

ii) $\varepsilon \in FRE$

iii) $a \in FRE, \forall a \in \Sigma$

iv) Scalar Multiplication: $(\lambda r) \in FRE, \forall \lambda \in L$ where $\lambda$ is a scalar and $r \in FRE$.

v) Addition: $(r_1 + r_2) \in FRE, \forall r_1, r_2 \in FRE$

vi) Concatenation: $r_1 r_2 \in FRE, \forall r_1, r_2 \in FRE$

vii) Kleene closure: $r^* \in FRE, \forall r \in FRE$

**Definition 2.7:** The Fuzzy language [3, 18] $\|r\|$ elucidated by fuzzy regular expression $r$ can be defined using the following rules:

i) $\|\emptyset\|(u) = 0, \forall u \in \Sigma^*$

ii) For $r \in \Sigma \cup \{\varepsilon\}, \|r\| = f_r$, where is $f_r$ the characteristic function of $r$ defined by

$$f_r(u) = \begin{cases} 1 & \text{if } a = r, \\ 0 & \text{otherwise}; \end{cases}$$

iii) $\|\lambda r\| = \lambda \otimes \|r\|\ \forall\ \lambda \in L$ and $r \in FRE$

iv) $\|r_1 + r_2\| = \|r_1\| \vee \|r_2\|,\ \forall\ r_1, r_2 \in FRE$

v) $\|r_1 r_2\| = \|r_1\|\ \|r_2\|,\ \forall\ r_1, r_2 \in FRE$

vi) $\|r^*\| = \|r\|^*\ \forall\ r_1, r_2 \in FRE$

**Definition 2.8:** Follow automata $A_f(r)$ [10] is a quotient of position automata $A_{pos}(r)$ and can be defined by the equivalence relation $\equiv_f$ over $pos(r)$ by

$i \equiv_f j$ iff (i) both $i, j$ or none belong to last $(r)$ and

(ii) follow $(r, i) =$ follow $(r, j)$

Using the above equation (i) and (ii), we can construct follow automata.

## 3. Fuzzy automata from fuzzy regular expressions using the follow automata

Let $L$ be an integral lattice-ordered monoid, and $r$ be a fuzzy regular expression over an input alphabet $\Sigma$. Consider a fuzzy regular expression $f_{re}$ over $\Sigma \cup Y$, where Y is a set consisting of scalar values which lies between 0 to 1. We first construct follow automata $A_f(f_{re}) = (Q_f, \Sigma \cup Y, \delta^{Q_f}, 0, \tau^{Q_f})$. Fuzzy automata can be constructed using the follow automata.

**Proposed Approach:** The proposed approach is represented in Fig. 2.

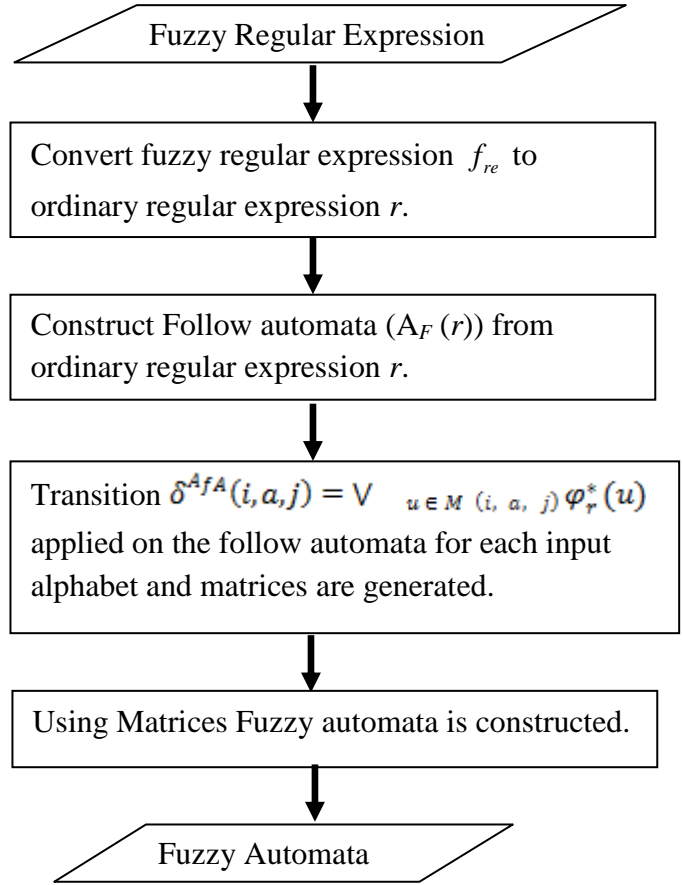

**Fig. 2**: Proposed approach for the conversion of Fuzzy regular expressions to Fuzzy Automata

**Example 3.1** Consider the fuzzy regular expression [3] $f_{re} = 0.2((0.1(ab)^*)^* + b)$ over the input alphabet $\{a, b\}$. Now convert $f_{re}$ into a regular expression $r = \lambda((\mu(ab)^*)^* + b)$ over the input alphabet $\{a, b, \lambda, \mu\}$ where $\Sigma = \{a, b\}$ and $Y = \{\lambda, \mu\}$. The marked regular expression is $r' = \lambda_1((\mu_2(a_3 b_4)^*)^* + b_5)$, and the homomorphism mapping $\varphi_r$ is given by

$$\varphi_r = \begin{pmatrix} a & b & \lambda & \mu \\ 1 & 1 & 0.2 & 0.1 \end{pmatrix}$$

Fig. 3 depicts the follow automata $A_f(r)$. Green color represents the final states in Fig. 3.

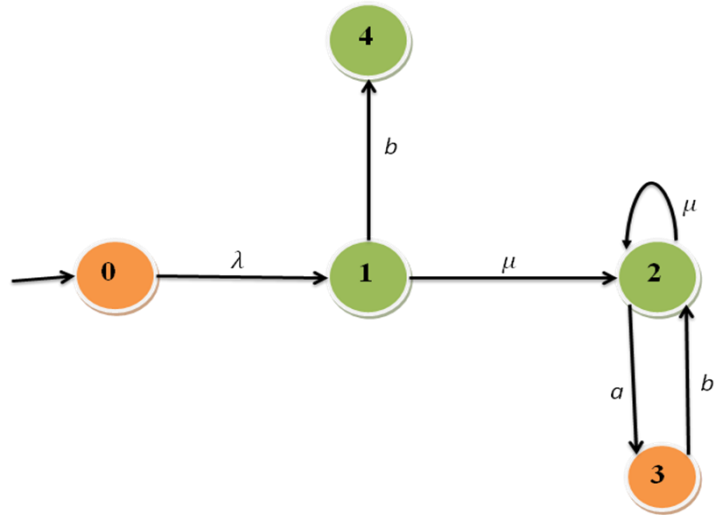

**Fig. 3**: Follow automata $A_F(r)$.

For any random $a \in \Sigma$ and $i, j \in Q_f$, define sets $P(i, a, j)$ and $M(i, a, j)$ are defined by Stamenkovic and Ciric [3] as follows:

$$P(i, a, j) = \{u \in U_y(a) \mid \delta^{AfA}(i, a, j) = 1\} \quad \ldots\ldots(1)$$

$$M(i, a, j) = \{u \in P(i, a, j) \mid u \text{ is minimal in } P(i, a, j) \text{ with respect to } \leq_{em}\} \quad \ldots\ldots(2)$$

Stamenkovic and Ciric [3] defined the transition using the following equation:

$$\delta^{AfA}(i, a, j) = \bigvee_{u \in M(i, a, j)} \varphi_r^*(u) \quad \ldots\ldots(3)$$

With the help of these equations, we can find out the minimum value in the path with respect to input alphabet.

For input alphabet, $a$, we will find out the minimum path, and if $\delta^{AfA}(i, a, j) = 0$, that means there is no input alphabet in that path.

$$\delta^{A_{fA}}(0, a, 3) = \varphi_a^*(\lambda\mu a) = 0.3 \otimes 02 \otimes 1 = 0.1,$$
$$\delta^{A_{fA}}(1, a, 3) = 0.2,$$
$$\delta^{A_{fA}}(2, a, 3) = 1,$$

For $(i, j) \notin \{(0, 3), (1, 3), (2, 3)\}$

For input alphabet $b$, we will find out the minimum path.

$$\delta^{A_{fA}}(0, b, 4) = 0.2,$$
$$\delta^{A_{fA}}(1, b, 4) = 1,$$
$$\delta^{A_{fA}}(3, b, 2) = 0.1,$$

If $\delta^{A_{fA}}(i, b, j) = 0$, means that there is no input alphabet $b$ in this path.

For $(i, j) \notin \{(0, 4), (1, 4), (3, 2)\}$

For input alphabet $a$, with the help of fig. 3, we can find,

$$M(0, a, 3) = \{\lambda\mu a\},$$
$$M(1, a, 3) = \{\mu a\},$$
$$M(2, a, 3) = \{x\},$$
$$M(i, a, j) = \emptyset \text{ for all other case.}$$

For input alphabet $a$, with the help of fig. 3, we can find,

$$M(0, b, 4) = \{\lambda b\},$$
$$M(1, b, 4) = \{b\},$$
$$M(3, b, 2) = \{\mu b\},$$
$$M(i, a, j) = \emptyset \text{ for all other case.}$$

The fuzzy transition relations $\delta_a^{A_{fA}}$, $\delta_b^{A_{fA}}$ and the fuzzy set of the final state $\tau^{A_{fA}}$ of the fuzzy automaton $A_{fA}(r)$ are given by,

$$\delta_a^{A_{fA}} = \begin{bmatrix} 0 & 0 & 0 & 0.1 & 0 \\ 0 & 0 & 0 & 0.1 & 0 \\ 0 & 0 & 0 & 1 & 0 \\ 0 & 0 & 0 & 0 & 0 \\ 0 & 0 & 0 & 0 & 0 \end{bmatrix}, \quad \delta_b^{A_{fA}} = \begin{bmatrix} 0 & 0 & 0 & 0 & 0.2 \\ 0 & 0 & 0 & 0 & 1 \\ 0 & 0 & 0 & 0 & 0 \\ 0 & 0 & 0.1 & 0 & 0 \\ 0 & 0 & 0 & 0 & 0 \end{bmatrix}, \quad \tau^{A_{fA}} = \begin{bmatrix} 0.2 \\ 1 \\ 1 \\ 0 \\ 1 \end{bmatrix}$$

With the help of these matrices, fuzzy automaton can be constructed as follows:

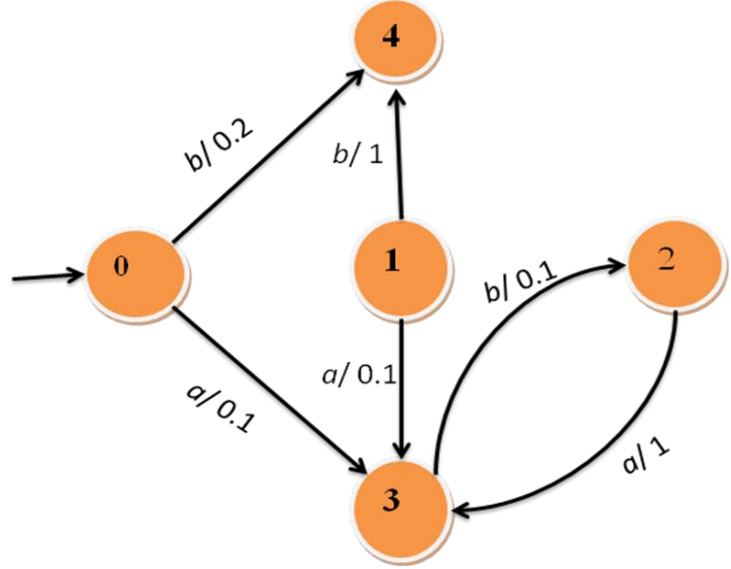

**Fig.4:** Fuzzy automata $A_{fA}$ ( $f_{re}$ )

Fig. 4 represents the efficient Fuzzy finite automaton of the fuzzy regular expression $r = 0.2((0.1(ab)^*)^* + b)$ over the input alphabet.

**Theorem 3.1.** Given L be an integral lattice–ordered monoid and $r$ be a fuzzy regular expression over the input alphabet $\Sigma \cup Y$ representing the language $\|r\|$. Let $A_{fA}$ be fuzzy automata obtained using the follow automata, then $L(A_{fA}) = \|r\|$.

**Proof.** First for $\varepsilon$ -transition, we have $L(A_{fA})(\varepsilon) = \tau^Q(q_0) = \tau^Q(q_0) = \|r\|(\varepsilon)$.

Then, for every string $s \in \Sigma^+$, where $s = u_1 u_2 \ldots \ldots u_n$, and all $u_1, u_2, \ldots, u_n \in \Sigma$,

$\|r\|(s) = L(A_{fA})(s) = (\delta^Q_{u_0} \circ \ldots \ldots \circ \delta^Q_{u_n} \circ \tau^Q)(q_0),$

$= \bigvee_{q_0 \ldots q_n \in Q} (\delta^Q_{u_1})(q_0, q_1) \otimes \ldots \ldots \otimes (\delta^Q_{u_n})(q_{n-1}, q_n) \otimes (\tau^Q)(q_0)$

$= (\delta^Q_{u_1} \circ \ldots \ldots \circ \delta^Q_{u_n} \circ \tau^Q)(q_0)$

$= L(A_{fA}).$

## 4. Conclusion and Future Scope

In this paper, the concept of follow automata has been extended for the metamorphosis of fuzzy regular expressions to fuzzy automata. Follow automata is a quotient of position automata. Compared with the existing approaches in the literature, the proposed approach will generate a lesser number of states without increasing the time complexity.